\documentclass[twocolumn,journal]{IEEEtran}
\usepackage{graphicx}
\usepackage{amsmath}
\usepackage{amssymb}
\usepackage{comment}

\usepackage{amsthm}
\usepackage{cite}
\usepackage{float}
\usepackage{color}
\usepackage{balance}

\begin{document}
\title{Secure Communications in the Presence of \\ a Buffer-Aided Wireless-Powered Relay \\ With Self-Energy Recycling}
\author{Ahmed El Shafie, \emph{Student Member, IEEE}, and Naofal Al-Dhahir, \emph{Fellow, IEEE}
\thanks{The authors are with the Electrical Engineering Dept., University of Texas at Dallas, USA.}
\thanks{This paper was made possible by NPRP grant number 6-149-2-058 from the Qatar National Research Fund (a member of Qatar Foundation). The statements made herein are solely the responsibility of the authors.}}
\date{}
\maketitle
\begin{abstract}
We consider a relay-assisted wireless network, where the energy-harvesting buffer-aided relay node is powered by radio-frequency signals from a source node wishing to communicate with its destination. We propose two secure cooperative protocols for a network composed of a source node equipped with a data buffer communicating with its destination in the presence of a buffer-aided relay node and an eavesdropper. Our proposed protocols are designed based on the channel state information and the buffer state information at the source and relay nodes. The burstiness of data at the source node queue (buffer) and the energy recycling process at the relay are taken into account in our analysis. In addition, we take the decoding and signal processing circuit power consumption constraints into consideration.
For the limiting case of an \emph{infinite-size} battery at the relay, we derive a sufficient condition for the energy queue to saturate.
Our numerical results demonstrate the throughput gains of our proposed protocols.
\end{abstract}
\vspace{-0.4cm}
\section{Introduction}
\vspace{-0.1cm}
Radio-frequency (RF) energy harvesting represents the capability of converting the received RF
transmissions into direct current (DC) electricity. Wireless-powered relaying (WPR), in which information is transmitted from the source to both the destination
and an energy-constrained relay node that is powered by RF energy signals, is one of the attractive applications of simultaneous wireless information and power transfer (SWIPT). Most of the existing research work on WPR assumes half-duplex relaying and adopts either time-switching-based relaying (TSR) \cite{reference5} or power-splitting-based relaying (PSR) schemes \cite{reference5}. In \cite{reference9}, the authors investigate a time-switching-based full-duplex WPR system, where the energy-harvesting relay node operates in a full-duplex
mode with simultaneous information reception and transmission.

Unlike the existing TSR or PSR protocols, which are the most commonly studied schemes for SWIPT in the literature, we propose simpler protocols which do not require a strict synchronization time switching process as in TSR or require a dedicated power splitting circuit that increases the complexity of the hardware design as in PSR. In addition, TSR and PSR require the optimization over the power-splitting ratio and switching-time, which consumes energy and complicates system design. The investigated full-duplex system in \cite{reference9} suffers from severe self-interference and it requires additional energy consumption at the relay node to implement the sophisticated analog and/or
digital interference-cancelation scheme \cite{reference10}. In our proposed protocols, the self-interfering channel at the
relay node is beneficial since it enables the relay to reuse its own transmitted energy. This actually represents the so-called self-energy recycling (loop
energy) \cite{reference11}, i.e., the energy used for information transmission can be harvested by the relay and reused in addition to
the RF energy harvested from the ambient transmissions. In \cite{reference11}, the authors proposed an interesting idea for energy recycling through the self-interfering link. The relay adopts a recycling energy protocol in which \emph{half} of the time slot is used for data transmission from the source to the relay while the second \emph{half} of the time slot is used for simultaneous data transmission from the relay to the destination and energy transfer from the source to the relay.
In the second transmission phase, the relay recycles its own transmit energy.

In this letter, we explore the potential benefits of channel state information (CSI), the buffering capability of the relay node, the burstiness of source data (i.e. sporadic and randomness of the data arrivals at the source node), the presence of a direct link between the source and its intended destination, and the buffer state information at various system nodes to further enhance the throughput of the system studied in \cite{reference11}. Furthermore, we assume the presence of an eavesdropper and impose hardware constraints such as \emph{limited-power transmissions}, a \emph{fixed-modulation scheme} and \emph{limited processing and decoding energy consumption}.

We consider a buffer-aided relay channel where a source communicates with its destination in the presence of an energy-constrained buffer-aided relay node and a potential eavesdropper. The contributions of this letter are summarized as follows. We design two protocols for secure communications between a source and its destination based on the CSI and the buffer state information (i.e. dynamics of the system queues). Unlike \cite{reference9,reference11}, we assume the presence of a direct link between the source and its destination and investigate the presence of an eavesdropping node in the system. Furthermore, we include the energy consumption due to data and decoding processing in our analysis. Unlike most existing work, we assume that the relay energy and data queues are limited in size. Moreover, we consider energy harvesting recycling where the energy-constrained relay can recycle the energy transmitted from its transmit RF chain to its RF-to-DC conversion circuits through the loopback channel. For the limiting case of an \emph{infinite-size} battery at the relay, we derive a \emph{sufficient condition} for the energy queue to \emph{saturate}.
\vspace{-0.3cm}
\section{System Model and Assumptions}
\vspace{-0.1cm}
We consider the randomize-and-forward relay channel \cite{mo2012relay}, where a source, $S$, communicates with its destination, $D$, with the assistance of an energy-constrained relay, $R$, and in the presence of an eavesdropper, $E$. In the randomize-and-forward relaying protocol, the
source and relay nodes use different codebooks to transmit the secret
messages \cite{mo2012relay} which increases the security of the system and increases ambiguity at the eavesdropper. We make the following assumptions in this letter. The relay is assumed to be an energy harvester that is solely powered by the RF signals from the source, $S$ \cite{reference11}. It is equipped with two antennas while all other nodes are each equipped with one antenna. Time is partitioned into slots each with a duration of $T=1$ seconds.\footnote{Since the time slot duration is normalized, the power and energy terms are used interchangeably.} The channel has a bandwidth of $W$~Hz. The source node, $S$, has an unlimited-size data buffer (queue) to store its incoming traffic, denoted by $Q_S$. The relay maintains two limited-size buffers: an energy buffer (rechargeable battery) to store the energy transferred from the source to the relay, denoted by $Q_{\rm e}$, and a data buffer to store the relayed source data packets, denoted by $Q_R$. The maximum buffer size of $Q_R$ and $Q_{\rm e}$ are $\mathcal{C}_R$ packets and $\mathbb{E}_{\max}$ energy units, respectively. We assume a fixed-rate data transmission as in, e.g., \cite{sadek}. If a node transmits in a given time slot, it sends exactly one data packet of size $\mathcal{B}$ bits over the entire time slot. Hence, the spectral efficiency is $\mathcal{R}=\mathcal{B}/W/T$ bits/sec/Hz. We assume \emph{fixed-power} transmissions as in, e.g., \cite{sadek}. Without loss of generality, we assume that the source uses the same power level for both information and energy signal transmissions. The source and relay average transmit powers are $P_S$ Watts and $P_R$ Watts, respectively. Thermal noise is modeled as an additive white Gaussian noise (AWGN) random process with zero mean and variance $\kappa$ Watts/Hz. The relay cannot transmit and receive {\bf information} at the same time and over the same frequency band, which is a practical assumption because practical implementations of full-duplex schemes \cite{reference10} are typically costly and can suffer from a significant level of loop interference.

Each link experiences flat block-fading where the channel coefficient is assumed to be constant during a time slot duration and changes from one time slot to another identically and independently in a random manner. The channel coefficient of the $i-j$ link is denoted by $h_{i,j}$ and its gain, which is the squared magnitude of the channel coefficient, is denoted by $g_{i,j}=|h_{i,j}|^2$, where $|x|$ denotes the absolute value of $x$.\footnote{To simplify notation, we omit the time index from the symbols.} It is assumed
that all the channel coefficients are known to all the nodes through feedback or channel estimation\footnote{The eavesdropper is assumed to be another non-hostile source node that utilizes the spectrum to communicate with the destination and the relay. CSI knowledge at the nodes is beneficial in our proposed protocols for: 1) efficient utilization of the secured time slots by the source and relay nodes since only one node is selected for data transmission in a given time slot, 2) preserving the node energy given that the energy at the relay node is limited, 3) avoiding the retransmission of the same data by the same node which reduces the system secrecy since the eavesdropper can then combine the data received at its receiver. If the eavesdropper's CSI is not known at the legitimate nodes, the proposed protocols can operate based on the link capacity and buffer states.}~\cite{reference11}. The energy dissipated in operating the transmit RF chain, i.e., the signal processing power consumption in the relay's transmit circuit caused by filters, frequency synthesizer, etc, is $\mathbb{E}_{\rm p} T=\mathbb{E}_{\rm p}$ joules per time slot. For the relay to transmit a data packet, it needs $\mathbb{E}_{\rm p}$ energy units to operate its circuits for \emph{one} second and $P_R$ energy units for its data transmission. Hence, the energy queue must maintain at \emph{least} $\mathbb{E}_{\rm t}=P_R+\mathbb{E}_{\rm p}$ energy units for the relay to transmit a data packet in a given time slot. In addition, we assume a certain processing and decoding energy level, denoted by $\mathbb{E}_{\rm d}$, which represents the needed energy to perform decoding and data processing at the relay when it receives data. This energy amount is a function of the number of received bits and other hardware constraints. We assume that the energy needed to decode a packet is $\mathbb{E}_{\rm d}$ energy units per time slot. The source uses a known time-invariant signal for powering the relay. This signal is known at all nodes including the eavesdropper; hence, its interference can be canceled out before information decoding.

For given channel realizations, the secrecy capacity of the $i-j$ link in the presence of eavesdropper $\ell$ is \cite{krikidis2009relay,mo2012relay}
 \begin{equation} \small
\small \begin{split} \small
C_{i,j}^\ell=\left[ \log_2\left(1+\frac{P_i g_{i,j}}{\kappa W}\right)-\log_2\left(1+\frac{P_i g_{i,\ell}}{\kappa W}\right)\right]^+
     \end{split}
\end{equation}
where $[\cdot]^+=\max\{\cdot,0\}$. If $g_{i,j}>g_{i,\ell}$, the secrecy capacity is greater than zero and a nonzero data rate is achievable. Otherwise, the link is unsecured and the achievable secrecy rate is \emph{zero}. The link is said to be in secrecy outage, if the instantaneous channel secrecy capacity is lower than the transmission rate. Since the transmission rate is $\mathcal{R}$, the secrecy outage condition is $C_{i,j}^\ell< \mathcal{R}$.

The energy harvested at the relay when it uses only one antenna, say antenna $r\in\{1,2\}$, and ignoring the negligible energy from the receiver noise (as in, e.g., \cite{reference9,reference11}) is
  \begin{equation} \small
\small \begin{split} \small
\label{eqn3}
\mathbb{E}_{{\rm H},1}=\eta \left(P_S T g_{S,R_r}+\delta P_R T g_{R_r,R_y}\right)
     \end{split}
\end{equation}
where $\delta P_R T g_{R_r,R_y}$ is the amount of self energy, $\delta=1$ if the relay is transmitting and zero otherwise, $r\ne y \in\{1,2\}$, and $0\le \eta\le 1$ is the RF-to-DC conversion efficiency. If the relay uses its two antennas to harvest energy from the source transmission, the total harvested energy is given by
  \begin{equation} \small
\small \begin{split} \small
\label{eqn4}
\mathbb{E}_{{\rm H},2}=\eta P_S T \left(g_{S,R_1}+g_{S,R_2} \right)
     \end{split}
\end{equation}
\vspace{-1cm}
\section{Proposed Protocols}
\vspace{-0.1cm}
In the following, we describe our proposed protocols which differ in terms of implementation complexity.
\subsubsection{Fixed-Antenna Assignment}
In this protocol, we assume that the antennas are always assigned to the same circuits. That is, one antenna is assigned for data reception and transmission and the other antenna is assigned for energy harvesting. When there is no data reception or transmission, both antennas are used for energy harvesting. Our proposed SWIPT with fixed-antenna assignment protocol is summarized as follows:
\begin{itemize}
\item If the direct link is secure and not in outage, i.e., $\mathcal{R}\le C_{S,D}^E$, the source transmits the packet at its queue head. The relay uses its two antennas for energy harvesting.
\item If the $S-D$ link is not secured or cannot support $\mathcal{R}$ bits/sec/Hz (i.e. $\mathcal{R}> C_{S,D}^E$), $\mathcal{R}\le C_{S,R}^E$, $Q_S$ is not empty, $Q_{R}$ is not full, and the energy queue $Q_{\rm e}$ has at least the energy needed for data decoding, i.e., $Q_{\rm e}\ge \mathbb{E}_{\rm d}$, $S$ sends data to $R$. In this case, the relay harvests energy from the source's RF transmission. Here, we assign priority to source transmission over relay transmission when $Q_S>0$, the relay data queue is not full, $Q_{\rm e}\ge \mathbb{E}_{\rm t}$, and the $R-D$ link securely supports $\mathcal{R}$ bits/sec/Hz. If the energy in $Q_{\rm e}$ is lower than the decoding energy, the relay can neither receive nor transmit data. If the relay has no data or/and its link cannot securely support $\mathcal{R}$ bits/sec/Hz or/and $Q_{\rm e}<\mathbb{E}_{\rm t}$, the source sends energy to the relay.
    \item If $\mathcal{R}\le C_{R,D}^E$, $\{\mathcal{R}> C_{S,D}^E,\mathcal{R}> C_{S,R}^E,Q_S>0\}$ or the source queue is empty, $Q_R$ is nonempty, and the relay has more than $\mathbb{E}_{\rm t}$ energy units, the relay transmits the packet at the head of $Q_R$ and the source sends energy to the relay by generating an energy~signal with average power $P_S$. A portion of the relay transmit energy loops back to the RF-to-DC conversion circuits and is further converted to DC electricity.
        \item If $\mathcal{R}>C_{S,D}^E$, $\mathcal{R}>C_{S,R}^E$, and $\mathcal{R}>C_{R,D}^E$, or if all the data queues are empty, the source sends energy to the relay if $Q_{\rm e}$ is not full. The relay uses its two antennas for energy harvesting. If $Q_{\rm e}$ is full, both the source and the relay remain completely idle.
\end{itemize}
From the above-mentioned cases, we have only \emph{six} possible states for the (source,relay) activities: (transmit information,receive information and energy), (transmit information,receive energy), (transmit information, idle), (transmit energy,simultaneously receive energy and transmit information), (transmit energy,receive energy), and (idle,idle). Hence, based on the CSI and the buffer state information, the control unit (which can be either the relay or the destination since they are the receiving nodes and they can feed back each other with all the necessary information) sends $3$ bits to inform all nodes about the current state.

Based on the protocol description above, the average service rate of the source data queue is given by
  \begin{equation} \small
\small \begin{split} \small
\mu_{\rm s}=\overline{\mathbb{P}_{S,D}}+\mathbb{P}_{S,D} \overline{\mathbb{P}_{S,R}} \Pr\{Q_{\rm e}\ge \mathbb{E}_{\rm d},Q_R<\mathcal{C}_R\}
     \end{split}
\end{equation}
where $\mathbb{P}_{i,j}$ is the probability of secrecy capacity outage of the $i-j$ link, and $\overline{\mathbb{P}_{i,j}}=1-\mathbb{P}_{i,j}$. Under the Rayleigh-fading channel model, the secrecy outage probability of the $i-j$ link in the presence of an eavesdropping node $\ell$ is given by \cite{4035982} $\mathbb{P}_{i,j}^\ell\!=\mathbb{P}_{i,j}=\!\Pr\{C_{i,j}^\ell< \mathcal{R}\}\!=\!1\!-\!\frac{\sigma_{i,j}}{\sigma_{i,j}+2^\mathcal{R} \sigma_{i,\ell}} \exp(-\kappa W\frac{2^\mathcal{R}-1}{P_{i}\sigma_{i,j}})$ where $\sigma_{i,j}$ is the average of $|h_{i,j}|^2$.

Since the relaying queue cannot transmit or receive more than \emph{one data packet} in a given time slot, its Markov chain is modeled as a \emph{birth-death} process. The transition probability from state $m$ to state $m+1$ at the relaying data queue, denoted by $\alpha_m$, is given by
    \begin{equation} \small
  \begin{split}
\alpha_m&=\alpha=\mathbb{P}_{S,D} \overline{\mathbb{P}_{S,R}}  \Pr\{Q_{\rm e}\ge \mathbb{E}_{\rm d},Q_S>0\}, \ 0\le m< \mathcal{C}_R
  \end{split}
\end{equation}
The transition probability from state $0<m\le \mathcal{C}_R$ to state $m-1$ at the relaying data queue, denoted by $\beta_m$, is given by
  \begin{equation} \small
\small \begin{split} \small
\beta_m&=\beta \!=\!  \overline{\mathbb{P}_{R,D}} \left(\mathcal{X} \Pr\{Q_S\!>\!0,Q_{\rm e}\!\ge\! \mathbb{E}_{\rm t}\}\!+\!\Pr\{Q_S\!=\!0,Q_{\rm e}\!\ge\! \mathbb{E}_{\rm t}\}\right), \\ & \,\,\,\,\,\,\,\,\,\,\,\,\,\,\,\,\,\,\,\,\,\,\,\,\,\,\,\,\,\,\,\,\,\,\,\,\,\,\,\,\,\,\,\,\,\,\,\,\,\,\,\,\,\,\,\ \forall m \in\{1,2,\dots,\mathcal{C}_R-1\},
\\
\beta_{\mathcal{C}_R} &\!=\!  \overline{\mathbb{P}_{R,D}} \left(\mathbb{P}_{S,D} \Pr\{Q_S\!>\!0,Q_{\rm e}\!\ge\! \mathbb{E}_{\rm t}\}\!+\!\Pr\{Q_S\!=\!0,Q_{\rm e}\!\ge\! \mathbb{E}_{\rm t}\}\right)
     \end{split}
\end{equation}
where $\mathcal{X}=\mathbb{P}_{S,D}\mathbb{P}_{S,R}$.

The average end-to-end secure throughput (i.e. the average number of successfully and securely received packets at the destination per time slot) is given by
  \begin{equation}
  \begin{split}
   \small
\small
\mu\!&=\!\underbrace{\overline{\mathbb{P}_{S,D}}\Pr\{Q_S\ne 0\}}_{\substack{{\hbox{\small successfully sent packets from}}\\ {\hbox{the source to the destination}}}}\!+\! \underbrace{\overline{\mathbb{P}_{R,D}} \left(\phi_1 +\phi_2\right)}_{\substack{{\hbox{\small successfully sent packets from}}\\{ \hbox{the relay to the destination}}}}
\end{split}
\end{equation}
where $\phi_1=\mathcal{X} \Pr\{Q_S\!>\!0,Q_{\rm e}\!\ge\! \mathbb{E}_{\rm t},0<Q_R< \mathcal{C}_R\}\!+\!\Pr\{Q_S\!=\!0,Q_{\rm e}\!\ge\! \mathbb{E}_{\rm t},0<Q_R< \mathcal{C}_R\}$ and $\phi_2=\mathbb{P}_{S,D}\Pr\{Q_S\!>\!0,Q_{\rm e}\!\ge\! \mathbb{E}_{\rm t},Q_R= \mathcal{C}_R\}\!+\!\Pr\{Q_S\!=\!0,Q_{\rm e}\!\ge\! \mathbb{E}_{\rm t},Q_R= \mathcal{C}_R\}$.
The energy queue has continuous service and arrival rates. Let $\mathbb{E}^\mathbb{T}_{\rm in}$ denote the amount of energy transferred from the source to the relay in time slot $\mathbb{T}\in\{1,2,\dots\}$ and $\mathbb{E}^\mathbb{T}_{\rm out}$ denote the amount of energy used by the relay in time slot $\mathbb{T}$. The energy queue evolves as follows:
  \begin{equation} \small
\small \begin{split} \small
Q_{\rm e}^{\mathbb{T}+1}=Q^{\mathbb{T}}_{\rm e}-\mathbb{E}^\mathbb{T}_{\rm out}+\mathbb{E}^\mathbb{T}_{\rm in}
     \end{split}
\end{equation}
where $Q_{\rm e}^{\mathbb{T}}$ denotes the amount of energy in $Q_{\rm e}$ at the beginning of time slot $\mathbb{T}$. In a given time slot, $\mathbb{E}^\mathbb{T}_{\rm in}$ is equal to zero, $\mathbb{E}_{{\rm H},1}$, or $\mathbb{E}_{{\rm H},2}$ based on the activity of the source (i.e. active or inactive) and the number of antennas used to collect the energy transferred from the source. If the relay decodes a packet, $\mathbb{E}^\mathbb{T}_{\rm out}=\mathbb{E}_{\rm d}$. If the relay transmits a packet, $\mathbb{E}^\mathbb{T}_{\rm out}=\mathbb{E}_{\rm t}$. If the relay is idle, $\mathbb{E}_{\rm out}^\mathbb{T}=0$.

Next, we derive a sufficient condition for the relay to be always saturated with energy when $\mathbb{E}_{\max}$ is very large, which represents a sufficient condition to consider the relay as a wireless node with reliable power supply. The \emph{minimum} average harvested energy throughout the network operation, which is obtained when only one antenna is always used for energy harvesting even when the relay uses its two antennas and assuming that there is no energy recycling, is $\eta P_S T \mathcal{E} \left[g_{S,R}\right]=\eta P_S \sigma_{S,R}$, where $T=1$ seconds and $\mathcal{E}\left[\cdot\right]$ represents the statistical expectation. According to Loynes' theorem, a queue is \emph{unstable} (i.e. \emph{overflows}) when its average arrival rate is higher than its average departure rate \cite{loynes1962stability}. Since the \emph{maximum} average transmit energy (i.e. departure rate) from the relay is $(P_R+\mathbb{E}_{\rm p}) \overline{\mathbb{P}_{R,D}}$ and the \emph{minimum} average transmit energy (i.e. arrival rate) is $\eta P_S \sigma_{S,R_r}$, the incoming energy at the energy queue is higher than the outgoing one when $\eta P_S \sigma_{S,R}> (P_R+\mathbb{E}_{\rm p})\overline{\mathbb{P}_{R,D}} \rightarrow P_S > \frac{(P_R+\mathbb{E}_{\rm p})\overline{\mathbb{P}_{R,D}}}{\eta \sigma_{S,R}}$, which represents a {\bf sufficient condition} for the energy queue to saturate with energy. In this case, the average service rate of the source queue can be rewritten as
  \begin{equation} \small
\small \begin{split} \small
\mu_{\rm s}\!=\!\overline{\mathbb{P}_{S,D}}\!+\!\mathbb{P}_{S,D} \overline{\mathbb{P}_{S,R}} \Pr\{Q_R<\mathcal{C}_R\}\le \underbrace{\overline{\mathbb{P}_{S,D}}+\mathbb{P}_{S,D} \overline{\mathbb{P}_{S,R}}}_{=\mu_{\max}}
     \end{split}
\end{equation}
with $P_S > \frac{(P_R+\mathbb{E}_{\rm p})\overline{\mathbb{P}_{R,D}}}{\eta \sigma_{S,R}}$.

Assume Bernoulli arrivals at the source with mean $\lambda_{\rm s}$ packets/slot as in \cite{sadek}. Then, according to Loynes' theorem, the
 \emph{sufficient condition} for the source queue to saturate is $\lambda_{\rm s}\ge \mu_{\max}$. If this condition is satisfied, the source will always be saturated with data packets. The transition probabilities of $Q_R$ are given~by
  \begin{equation} \small
  \begin{split}
  \alpha_m&=\mathbb{P}_{S,D} \overline{\mathbb{P}_{S,R}}, \ \beta_m =  \overline{\mathbb{P}_{R,D}}\mathbb{P}_{S,R}\mathbb{P}_{S,D}, \ \beta_{\mathcal{C}_R}=\overline{\mathbb{P}_{R,D}}\mathbb{P}_{S,D}
  \end{split}
\end{equation}
with $m< \mathcal{C}_R$.

Analyzing the Markov chain of the relaying queue, we get the following closed-form expressions for the local balance equations (derivation is omitted due to lack of space):
\begin{equation} \small
\label{balance}
\small
\small \Gamma_m \alpha_m=\Gamma_{m+1}\beta_{m+1}, 0\le m \le \mathcal{C}_R-1
\end{equation}
where $\Gamma_m$ denotes the probability of having $m$ packets at $Q_{R}$. Using the balance equations successively, the stationary distribution of $\Gamma_m$ for $Q_{R}$ occupancy is given by
\begin{equation} \small
\small
\small \Gamma_m=\Gamma_0 \prod_{n=0}^{m-1}\frac{\alpha_n}{\beta_{n+1}}, \hbox{where} \ \Gamma_0=\Big(1+\sum_{m=1}^{\mathcal{C}_R}\prod_{n=0}^{m-1} \frac{\alpha_n}{\beta_{n+1}}\Big)^{-1} \label{state_probabilities}
\end{equation}
where $\Gamma_0$ is obtained using the normalization condition $\sum_{m=0}^{\mathcal{C}_R} \Gamma_m=1$.
\subsubsection{Adaptive-Antenna Assignment}
 Next, we assume that the relay adopts an adaptive scheme for efficient antenna usage as in \cite{reference9}. That is, the two antennas can be used adaptively for data transmission and reception, and RF energy harvesting. Without loss of generality, we use a subscript $r\in\{1,2\}$ to indicate the relay's first and second antennas, respectively. If only one of two antennas at node $i$ is selected for data transmission, the maximum secrecy capacity of node $i$ when it communicates with node $j$ using the antenna with the highest link gain in the presence of eavesdropper $\ell$ is given by
$C_{i,j}^\ell=\max\{C_{i_1,j}^\ell,C_{i_2,j}^\ell\}$.
Similarly, if node $i$ is equipped with one antenna and node $j$ is equipped with two antennas and only one of them is selected in a given time slot for data reception, $C_{i,j}^\ell$ is given by $C_{i,j}^\ell=\max\{C_{i,j_1}^\ell,C_{i,j_2}^\ell\}$. Consequently, we conclude that having additional antennas increases the achievable secrecy rate. Since the nodes transmit their data at a fixed rate $\mathcal{R}$, we select the antenna which achieves the target rate and assign the other antenna for energy harvesting. Hence, the relay harvests more energy from the ongoing transmission if the antenna with the lowest channel gain to $S$ is selected for data transmission/reception.

As in the previous subsection, we can write down the expressions of $\mu_{\rm s}$, $\alpha_m$, $\beta_m$ and $\mu$. In this protocol, we have \emph{six} possible states for the activity of the source and the relay as in the fixed-antenna assignment protocol, but \emph{one} additional bit is required for the identification of the antenna index used at the relay in case of information reception/transmission. The total number of states under this protocol is $12$; hence, the control unit sends $4$ bits to inform the nodes about the current~state. \vspace{-0.4cm}
\section{Simulations and Concluding Remarks}
\vspace{-0.1cm}
In this section, we evaluate the performance of the proposed protocols using $50000$ time slots. For our numerical results, we assume the Rayleigh-fading channel model where each channel is distributed according to a circularly-symmetric Gaussian random variable with zero mean and unit variance, the AWGN power at a receiving node is normalized to $\kappa W=1$ Watts, and the energy needed for data decoding is $\mathbb{E}_{\rm d}=3$ joules. The time slot duration is normalized to $T=1$ seconds and $\mathbb{E}_{\rm p} T=\mathbb{E}_{\rm p}=2$ joules. Furthermore, the transmit powers of the source and the relay are $P_S=P_R=20$ Watts. In addition, the maximum buffer size of $Q_R$ is $\mathcal{C}_R=10$ packets and the maximum energy level at $Q_{\rm e}$ is $\mathbb{E}_{\max}=40$ joules. Note that the levels of the transmit powers and energies are large because the noise power and the time slot duration are normalized. The arrivals at the source data queue are assumed to be Bernoulli random variables with mean $\lambda_{\rm s}=1$ packets per time slot. Moreover, we assume that the conversion efficiency is $\eta=1$.

\begin{figure}
\vspace{-0.8cm}
\centering
\includegraphics[width=1\columnwidth]{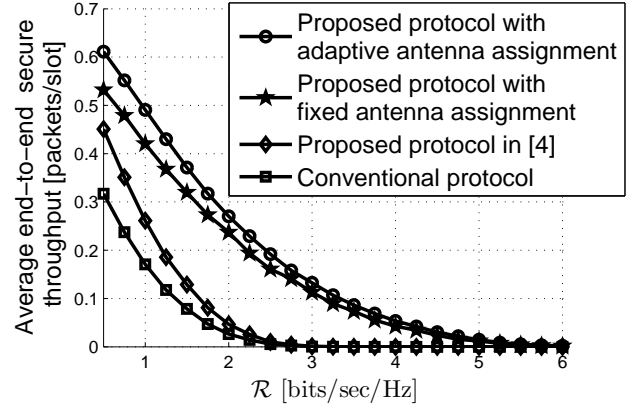}
\caption{Average end-to-end secure throughput of our proposed protocols compared to the protocol proposed in \cite{reference11}.}
\label{fig2}
\vspace{-0.5cm}
\end{figure}

In Fig. \ref{fig2}, the average end-to-end secure throughput, which represents the average number of successfully and securely received packets at the destination, in packets per time slot (packets/slot) is plotted against the target spectral efficiency, $\mathcal{R}$, for our
proposed protocols and compared to two baseline protocols, namely the protocol in \cite{reference11} for our randomize-and-forward buffer-aided relay system and what we call the conventional protocol where the relay node is not equipped with a data buffer and there is no energy recycling. We observe that significant
throughput gains are achieved by our proposed protocols for all values of $\mathcal{R}$ because our protocols take into consideration energy recycling, buffer state information at nodes, eavesdropper impact, and CSI, resulting in more efficient use of the available energy at the relay and the secured time slots. In addition, our adaptive-antenna assignment protocol outperforms the fixed-antenna assignment protocol, but at the expense of increased implementation complexity.
\bibliographystyle{IEEEtran}
\bibliography{IEEEabrv,term_bib}
\vspace{-0.2cm}%
\end{document}